# Architecture and Applications of IoT Devices in Socially Relevant Fields

Anush Lakshman S[1*], Akash S[2], Cynthia J[2], Gautam R[2], Ebenezer D[2]

[1*]Department of Mechanical Engineering, Iowa State University of Science and Technology, 2025, Ames, 50011, Iowa, USA.
[2*]Department of Mechanical Engineering, Sri Sivasubramaniya Nadar College of Engineering, Kalavakkam, 603110, Tamilnadu, India.

*Corresponding author(s). E-mail(s): anush18014@mech.ssn.edu.in;
Contributing authors: akash18010@mech.ssn.edu.in;
cynthia18036@mech.ssn.edu.in; gautam18045@mech.ssn.edu.in;
ebenezerd@ssn.edu.in;

**Abstract**

A multitude of IoT-enabled devices are continually being explored and introduced annually, fostering robust competition among researchers and businesses seeking to leverage the IoT landscape, given the substantial market potential of these devices. The selection of IoT architectures, communication protocols, and components is contingent upon the task's nature and the data sensitivity the device manages, and thus, evaluating their performance becomes crucial. This paper provides a comprehensive review of IoT-enabled devices, examining their architectures, communication protocols, and functionalities in socially significant domains such as healthcare, agriculture, firefighting, safety applications (including women's/individual safety, emergency alerts, etc.), home surveillance, and mapping – areas significantly impacting the general public. A comprehensive literature survey has been conducted to understand the various IoT architectures and compare the suitability of the aforementioned corresponding applications.

# 1 Introduction

The concept of electronic devices communicating across distances traces back to the early 1800s when Baron Shillings in Russia invented the electromagnetic telegraph, allowing machines to exchange electrical signals. With the advent of the Internet, this form of communication took on new dimensions and rapidly evolved. One of the first connected devices, employing a similar communication technology, emerged in the 1980s—a Coca-Cola vending machine modified by programmers at Carnegie Mellon University for object counting (i.e.) the number of soda cans left in the machine and transferring the information to a lab down the hallway. The concept of communicating information through connected devices was termed as 'embedded internet' or 'pervasive computing'. This idea of connected devices was later termed as 'Internet of Things (IoT)' by Kevin Ashton in 1999. The further development of this idea led to the introduction of WiFi-compatible refrigerators by LG in 2000. After this, the growth of IoT was substantial over the next decade, which saw the number of connected devices overtake the world population. In addition to many organizations performing active research in IoT, this surge attracted attention from global industrial leaders such as Apple, Samsung and Google to invest more in this technology, which led to significant strides of development in taken in the field.

By 2013 and 2014, IoT devices started incorporating more sensors to enhance performance, obtain and transfer more information. At the time, this addition was revolutionary, leading to significant advancements in various disciplines. For instance, Dublin transformed into the first smart city in 2014, which leveraged the use of IoT-based sensors and employed it to various utilities like smart bins, city sound monitoring systems, and flood level monitoring sensors. In the defense sector, IoT aided in battlefield information acquisition, processing, and providing alerts to warfighters. With the rapid advancement of health monitoring sensors, IoT also ventured into the medical industry around 2018, facilitating remote patient monitoring for healthcare professionals.

More recently, during the pandemic in 2020, IoT played a pivotal role in contact tracing across hospitals, workplaces, and crowded areas, as well as in tracking vaccine supplies. In addition to this, the rapid development in graphics processing units (GPUs) paved way for compute-intensive techniques like deep learning (DL) and generative modelling to be utilized more in various applications Sai et al (2024); Xu et al (2024). The substantial increase of IoT devices is predicted to continue as forecasts predict a staggering rise to approximately 70 billion devices by 2030 Sharma et al (2018). The overall timeline is illustrated in Fig 1, which is incomplete to account for the anticipated growth.

The most essential step in establishing a secure and functional framework, is selection of the architecture suitable for the field of application. In order to perform this task efficiently, we first need to understand the different types of architectural frameworks that are currently being utilized. The principles of the popularly employed frameworks are elaborated in Section 1.1.

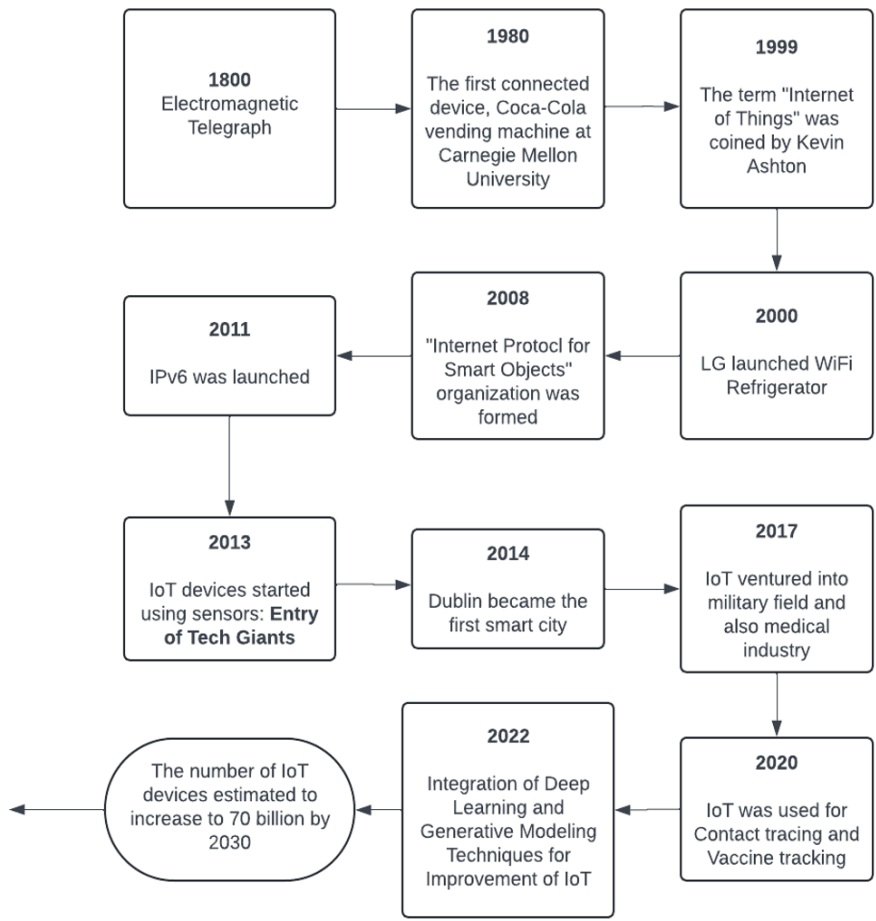

**Fig. 1**: The notable timelines of the evolution of IoT

## 1.1 IoT Architecture

There are three types of IoT architecture, namely, three-layer, four-layer, and five-layer architecture. The three-layer architecture consists of the perception layer, the network layer, and the application layer. As shown in Fig. 2a, the perception layer consists of physical hardware components like sensors, transmitters, and monitors, which are used to sense and transmit the data. The network layer bridges the perception and the application layer. This connection is established with the aid of technologies such as Wireless Fidelity networks (WiFi), Long Range Wide Area Networks (LoRaWAN), Global System for Mobile Communication (GSM), etc. The application layer refers to the applications where IoT has been used, like smart farming, smart cities, smart wearables, etc. The three-layer architecture simply transmits the data throughout the architecture and thus is more susceptible to threats and leakage of private data; hence, it can be used in applications where the data is not sensitive or confidential. However, to tackle such security issues, an additional layer, called the support layer, is included in the four-layer architecture as shown in Fig. 2b. The main use of this layer is to secure the data by methods such as password protection and encryption, thus restricting access to authenticated users only. The other responsibility of this layer is to transmit data back to the network layer via wired or wireless networks. The first two architectures are suitable for iterative work or data transmission but it doesn't attempt to improvise based on the data that it records. The five-layer architecture addresses it by another layer called the business layer, as shown in Fig. 2c. This layer acts as the center of management for the whole IoT network by managing the data that is private and public, determining the creation, storage, and changes to the information. This also establishes control over the profit models of the IoT. Here, the network layer is a combination of the transport and processing layer. The transport layer is simply used to transmit the data via wireless networks. The processing layer is used to filter out the data coming from the network layer, transmit only the required data to the next layer, and eliminate irrelevant information. Burhan et al (2018)

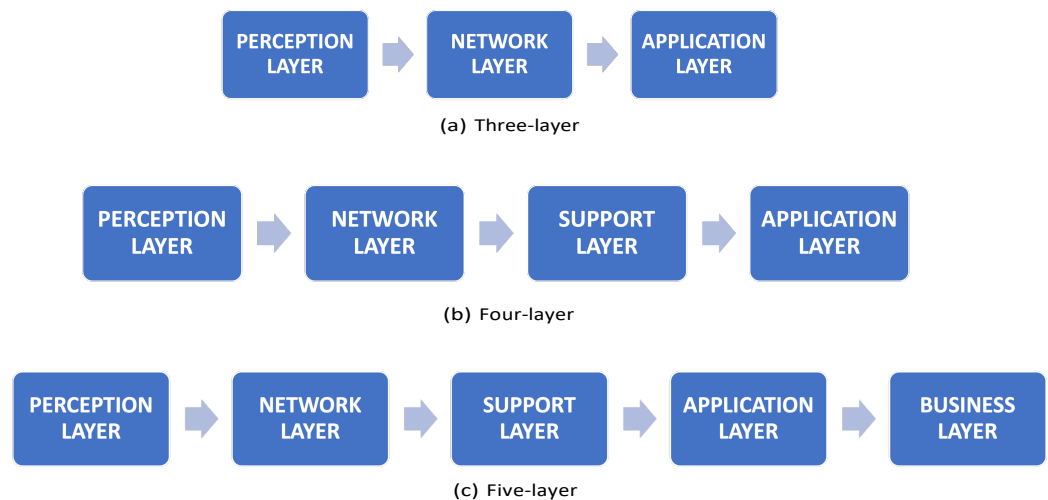

**Fig. 2**: IoT Architecture Classification

After selection of the architecture, the next step would involve construction of each layer in the architecture. The most important layer in an IoT framework is the communication layer or network layer, which establishes the connection between the sensing layer and the application layer. There are multiple methods through which this connection is established and these methods are referred to as communication protocols. This paper surveys the different architectures, layers and communication protocols that have been utilized in various applications such as healthcare, agriculture, firefighting, safety and danger alert, surveillance and mapping. The rest of the paper will be structured as follows:

- Discussion about the various communication protocols that can be employed as the network layer.
- The literature-based analysis to select a suitable architecture in fields such as healthcare, agriculture, firefighting, safety and danger alert, surveillance and mapping.
- Based on the surveyed literature, potential future work in each field analyzed by integration of technologies like Generative Modelling, Artificial Intelligence and Machine Learning.

## 1.2 IoT Communication Protocols

Communication protocols play a pivotal role in IoT networks as they enable information transfer from the perception layer to the application layer. The classification of these communication protocols can be done based on various aspects as illustrated in Fig 3. One classification method distinguishes them as Internet Protocols (IP) and non-Internet Protocols (non-IP), depending on their connection to the Internet via a gateway. IP examples include Wireless Local Area Network (WLAN), WiFi, and GSM, whereas non-IP examples consist of Bluetooth Low Energy Module (BLE), ZigBee, etc.

Secondly, on the basis of communication range, protocols are categorized as personal area network (PAN), local area network (LAN), and wide area network (WAN). PAN, like those used in wearables connecting to smartphones, covers a limited area (up to 30 meters). LAN, such as WiFi routers in homes, spans a larger range (about 100 meters) and transmits data at speeds up to 250 kbps but also supports a limited number of devices. WAN, typified by networks like RailwireWiFi in Indian railway stations, compromises speed (about 10 – 30 kbps) to cover larger areas. Dhilipkumar et al (2024).

Topology-wise, network protocols can be classified as star, ring, peer-to-peer (P2P), tree, bus, mesh, and hybrid. Star topology functions with a central hub (gateway) connecting end devices, ensuring reliability unless the hub fails, causing the entire network to fail. Ring topology involves devices connected in a circular manner, transmitting information unidirectionally or bidirectionally, but a failure in one device can disrupt the entire network. Tree topology employs routers connecting end devices to a gateway (root node), offering simplicity in error identification and maintenance. P2P allows direct communication between routers but lacks centralized file backup like star topology. Bus topology features a central cable connecting all nodes, offering cost-effectiveness but is prone to network failure if the main cable fails. Mesh topol- ogy connects source and destination via multiple pathways, either partially or fully, providing redundancy but requiring additional cabling and space. Hybrid topology combines multiple topologies, offering scalability but at higher costs. The pictorial representations of all these network topologies are illustrated in Fig 4.

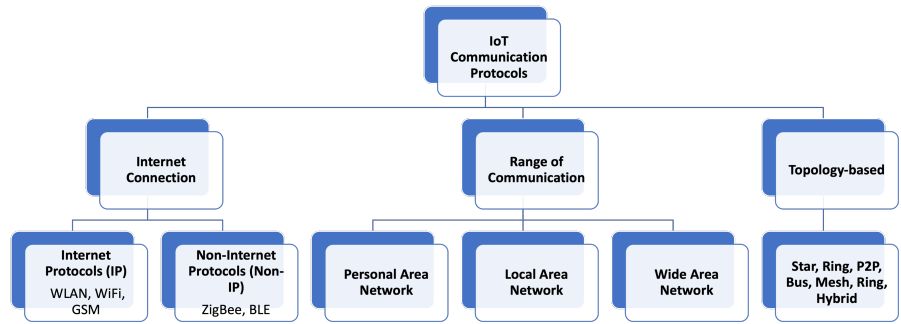

**Fig. 3**: Classification of IoT Communication Protocols

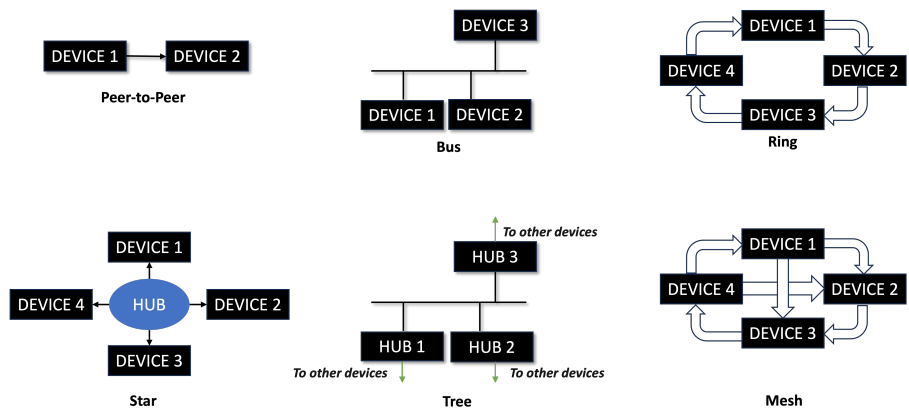

**Fig. 4**: Topology-based Classification of IoT Communication Protocols

## 2 Impact of IoT in Biological Fields

### 2.1 Healthcare

Medicine is arguably the largest sector that caters to the needs of everyone around the world through hospitals and clinics. The recent developments in this field is the inclusion of IoT into the healthcare sector to provide faster and more accurate diagnosis and also monitor the patient on a 24-hour basis Pinto et al (2017); Rejeb et al (2023); Nissar et al (2024). In order to execute these tasks, a number of diverse devices compile, interpret and communicate information about the patient in real-time, thereby enabling a large amount of data to be obtained, processed, and analyzed in different ways using a variety of sensors and communication protocols to interact with cloud storage. The information on the cloud storage is sequentially tabulated by software and extracted by a remote device with an internet connection Kodali et al (2015).

### 2.1.1 Health Monitoring

A survey made by the World Health Organisation (WHO) has predicted that the number of individuals with chronic diseases, such as Diabetes and Alzheimer's, among many others, will increase to about 66% of the population by 2030 Mdhaffar et al (2017). In a situation such as this, it is natural to expect some methodology to look after the individual around the clock.

A generic architecture for a Health monitoring system falls under the broad classification of the five-layer architecture mentioned in Section 1.1. In Health monitoring systems, in particular, it is split into IoT Device Layer, IoT Gateway Layer, IoT Service Platform Layer, and IoT Core Network Rajput and Gour (2016). The IoT Device Layer consists of sensors that support standards such as Zigbee, Z-Wave, ANT, and Wi-Fi. Zigbee, Z-Wave, and ANT define wireless communication protocol stacks that enable hardware to communicate by establishing standard rules for coexistence, data representation, signaling, authentication, and error detection, much like Bluetooth and Wi-Fi.

The sensors in the perception layer vary based on their intended application. The Gateway Layer connects the sensors securely to the IoT Administration Stage. The Service Platform gives perfect IoT administration reflections that can be utilized to connect to different applications. The Core Network is an IP-based system that acts as a correspondence network to connect the physical elements of the other stages, facilitating data trade between the layers.

Field (2002) have inferred that the use of a Remote Patient Monitoring System in ICUs has led to a reduction in costs while simultaneously providing round-the-clock monitoring of the patient. Malasinghe et al (2017) used this to propose that conventional medical sensors, such as a thermal sensor or even a pulse oximeter, can be replaced by cameras and smartphones. Such a development will make remote monitoring more popular and easier to implement.

Especially with the advent of the COVID-19 pandemic, the need for a remote patient monitoring system increased tremendously. Antonio et al (2021) proposed a remote monitoring system to monitor patients based on the National Early Warning Score 2 (NEWS-2). McDuff et al (2014) considered photoplethysmographic imaging techniques to monitor stress level of patients remotely. This method uses non-invasive camera color signals to process data regarding blood oxygen levels and respiration rate, among others.

Mdhaffar et al (2017) proposed an architecture for a healthcare system using the IoT for Healthcare (IoT4HC) framework, as shown in Fig 5, which can monitor and treat patients remotely. It can be seen that an all-round comprehensive set of data can be acquired from the various sensors and their health can be monitored in real time.

The miniaturization and reduction in cost of these sensors over the past couple of decades has led to the increased popularity of wearable health devices in the form of Fitbits, Smarthwatches and health trackers Paul et al (2024). Wearable Devices can be popularly categorized under two broad categories based on their application. They are systems for patient monitoring and systems sor motion tracking. Patient Monitoring was meant for individuals whose bodies cannot perform a certain function by themselves and need external intervention to facilitate them to go about their daily activities. It largely refers to the round-the-clock monitoring of a patient and reporting the readings to the concerned doctor in a timely manner. Whereas, Motion Tracking devices are a result of wearable devices becoming popular even among the general public. It deals with devices that are commercially available to help individuals monitor their own fitness and exercise regimes and even sleep patterns. The individuals are directly responsible for accessing their own health data and there is no physician involved in the process. Some other such devices are discussed below in detail.

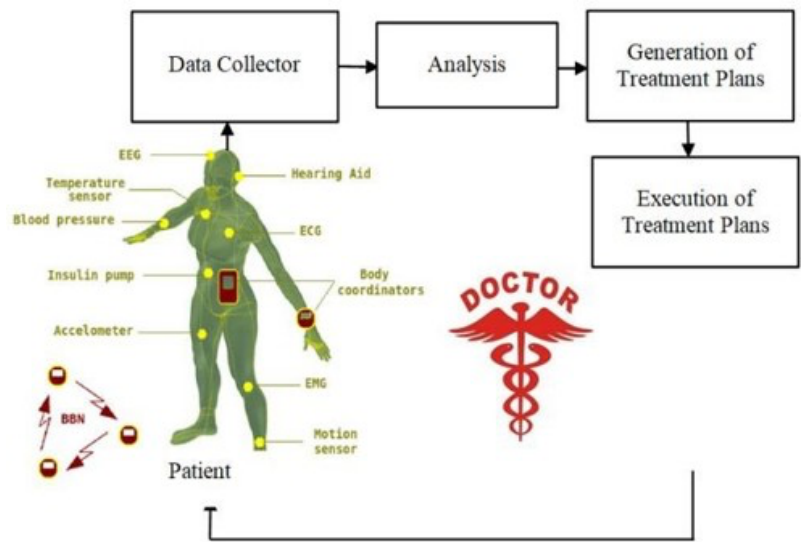

**Fig. 5**: IoT4HC Architecture Mdhaffar et al (2017)

1. **Systems for Patient Monitoring:** Wearable Devices like a watch or a patch can monitor an individual's signals in real time and make decisions accordingly. IoT-induced wearable device is an Insulin pump that monitors the sugar level in the blood and pumps the required dosage of insulin accordingly. Bergenstal et al (2010) implemented this continuous monitoring of blood sugar level using a Min- iMed Paradigm REAL-Time System (Medtronic) and performed experimentation to establish the superiority of continuous monitoring systems. Another sophis- ticated IoT-induced instrument, the Cardiac Rhythm Medical Device (CRMD), commonly known as the Pacemaker, besides controlling the heartbeat rate, also sends signals that can be used to detect heart seizures in advance.However, Stachel et al (2013) have concluded after extensive in-vitro experimentation that Radio Frequency (RF) emitters cause significant interference to the working of the CRMD.

Other commonly used devices are Hearing Aids and Bionic Arms Lili et al (2019). Adeel et al (2020) hold firm in their belief that integration of IoT in hearing aids is the way to the future although they demand high data rate, low latency, low computational complexity, and high security. Johansen et al (2017) explored the possibilities of using such a hearing aid as a customizable hearing device that can provide intelligent control of sound, by increasing and reducing loudness and speed of speech, as necessary. Lili et al (2019) developed the control system for an intelligent bionic manipulator that can save routines and also react according to the environment. The smart glove is another useful device which is used to aid blind people go about their daily activities. Gopi and Dr.Punarselvam (2019) developed a smart glove lined with ultrasonic and vibration sensors along with an inbuilt GPS sensor. While the GPS sensor gives the person global information, the other two sensors alert the individual of imminent threats in his surroundings. Vasanth et al (2019) developed an aid that can help people who are simultaneously partially deaf and blind by amplifying audio and converting text with the aid of the Google Speech API module.

Other wearable devices like Body Worn Smart Clothing and Wire based devices consists of an assortment of sensors that detect different vitals which can be used for patient monitoring. The commonly used sensors used to sense our vitals are body temperature, Electroencephalogram (EEG)/ Electrocardiogram (ECG), Blood Oxygen saturation level (SpO2), Blood pressure, Blood Sugar, Breathing CO, Alcohol, Position, Body Local Angle, Weight and some movement sensors Zhao et al (2011). The EEG/ECG sensors measure neural activity by measuring electrical potential across the scalp. The SpO2 sensor is a noninvasive sensor that measures the percentage of blood enriched with oxygen.

Fangyu et al (2021) proposed a framework to monitor the respiration of home quarantined COVID patients. The existing Wi-Fi signals in the room are used for analysis and the shortness of breath is estimated with frequency variation methods and statistical calculations. Similarly, Zhang et al (2019) proposed a breath detection system using Wi-Fi for indoor patients. This was later adapted by various other developers for various applications. Adib et al (2015) proposed the integration of smart homes with patient monitoring systems and emergency alert system in case any individual in the house falls sick.

The SmartPill is a device now commonly used in hospitals across the globe for gastric monitoring Hasler (2014). It consists of Pressure and pH sensors and a transmitter. So, even as the pill goes down an individual's oesophagus, the doctor can monitor the passage. This is an example of a patient monitoring system where the hardware is ingested into the patient. With further modification, it can prove to be a boon to people with chronic illnesses who need focused 24-hour monitoring.

With the increasing trust in AI and Machine Learning, the utilization of these techniques in medical diagnosis has also been done. For instance, Paul et al (2024) performed predictive health monitoring by use of data obtained from a Fitbit. Ramkumar et al (2023) utilized advanced Deep Learning Techniques such as Long-Short Term Memory network(LSTM) and Recurrent Neural Networks (RNNs) for predictive health monitoring of heart health of data obtained from smart health monitoring devices. Due to the increased application of AI, the base architectures are slightly modified to incorporate some functionalities required to incorporate AI. This can be seen in McDuff et al (2014), Antonio et al (2021), Ramkumar et al (2023) and Paul et al (2024).

2. **Systems for Motion Tracking:** The second set of devices comprise motion tracking devices available in the open market. Some of them are listed below Haghi et al (2017).

   - **Fitbit Flex -** Counts the steps and measures the quality of steps. It is small in size and is wrist-worn
   - **Withings Pulse -** Counts the steps and distance travelled. It records sleep time. It can also measure heart rate and the percentage of optimal sleep hours
   - **Misfit Shine -** Counts the steps, distance and daily calories burnt.It can double as a sleep tracker monitor and can even identify hours of light as well as deep sleep
   - **Jawbone -** Tracks user's sleep data, eating habits, calories burned, and daily activity. It also counts the steps and distance travelled.

   The representation of where these devices are worn is illustrated in Fig 6

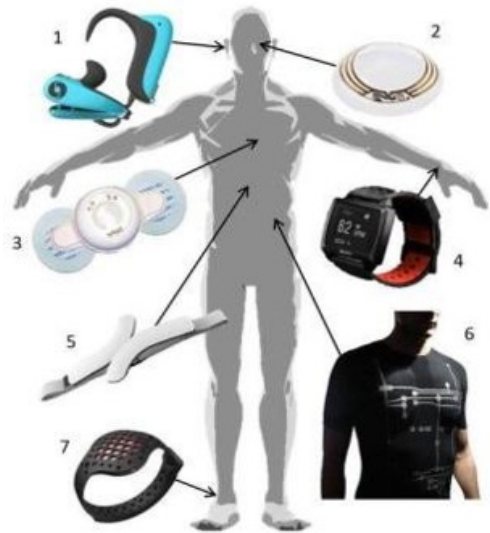

**Fig. 6**: Representation of Wearable Health Devices Dias and Paulo Silva Cunha (2018)

In common, all such wearable devices contain an assortment of ECG sensors, ambient light sensors, skin response sensors, bio-impedance sensors, barometric altimeters, accelerometers, compasses and gyroscopes. The barometric altimeter is used to track altitude by measuring the atmospheric pressure. Skin response sensors give an estimate of how much a user perspires. The bioimpedance sensor is capable of measuring the body tissues' resistance to electric currents. This feature allows a fitness device to capture different physiological signals, including respiration rate, heart rate, and galvanic skin response. Ambient light sensors can detect the amount of light in the environment. Using frameworks like the ones discussed in Section 2.1.1, the data from these sensors can be made available on the user's smartphone for direct access. The final layer, or the user interface, has many commercially available options, as discussed below.

Strava is a fitness app frequented by athletes and cyclists. FoodSpex is a comprehensive database of the number of calories each food contains. The BMI Calculator measures the Body Mass Index of the user. And Garmin is an app that offers the tools to help developers make apps for wearable devices. Fig 4 portrays a few other wearable health devices commonly available on the market along with the location of the body on which they are used to measure the vitals Dias and Paulo Silva Cunha (2018). They are (1) SensoTRACK ear sensor; (2) Google Contact Lens; (3) BioPatch$^{TM}$; (4) Smartwatch Basis PEAK$^{TM}$; (5) QardioCore; (6) Vital Jacket$^®$ t-shirt; and (7) Moov (activity tracker) respectively. All these devices combine the five layers into an easily accessible form. Fig 7 gives a general classification on the contributions of IoT in Healthcare and also represents the flow of signals in the system. The primary classification is based on Patient Monitoring Systems and Motion Tracking Systems. The presence of a doctor reflects the difference between the two as a mediator between data and the treatment in Patient Monitoring Systems. Jayatilleka and Halgamuge (2020) comprehensively classified a list of noteworthy contributions by IoT to Healthcare.

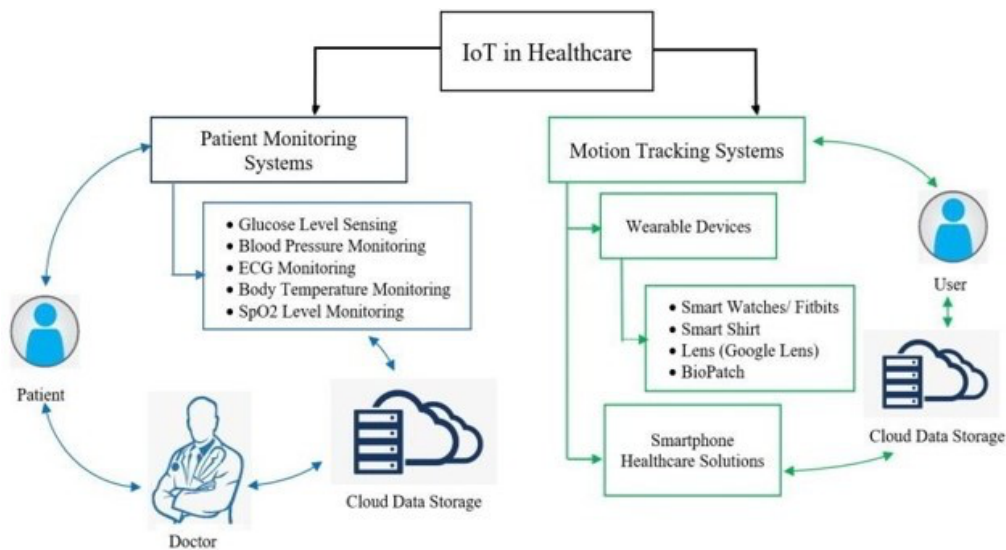

**Fig. 7**: Classification of IoT in Healthcare

### 2.1.2 Extended Applications of Unmanned Aerial Vehicles and Other Devices in Healthcare

Telehealth is a term used to represent all forms of administering treatment using electronic communication methods Tuckson et al (2017). It encompasses new and unique methods to detect diseases early and treat them consistently in a remote fashion.

Unmanned Aerial Vehicles (UAVs) have been around for a decade, and they are being integrated with IoT to a satisfactory level. With recent developments in this field in consideration. They can be extended to healthcare. The Amazon Prime Air is a UAV used by Amazon to deliver packages. The following developers have adopted the same UAVs to deliver medicine among other purposes.

Kumar et al (2019) developed a prototype Aero Ambulance Quadcopter that served as a medium to quickly transport to the hospital without waiting for traffic. It provides a low-cost, simple system of an Arduino UNO board, GPS sensor, Electronic Speed Controller (ECS), and a Brushless DC Motor (BLDC Motor). Similarly, in Rwanada, Africa, Zipline's Drones cover a range of up to 75 km transporting blood and other medications once an order has been placed for the same through WhatsApp Ackerman and Strickland (2018). Krishna et al (2018) proposed a model for Drone Ambulance. This proposed to send a drone packed with ECG sensor ahead of the ambulance for faster response times. The drone used Zigbee technology to communicate with the ambulance in advance so that the doctors could be prepared.

Sanjana and Prathilothamai (2020) proposed a drone that quickly delivered a first aid kit to an affected person. Another more ambitious approach is a drone that delivers Automated External Defibrillators (AEDs) which was developed by Rosamond et al (2020). It proposes that any individual in the vicinity with the minimum operating knowledge can use it to save a person's life without having to wait for an ambulance. With street monitoring systems the following types of UAVs can directly be deployed by the AI to save dozens of lives. But a system cannot operate alone without support. For such UAV-based delivery systems to succeed, it is necessary to develop good mapping techniques to direct the UAVs.

Moodables are mood-enhancing devices that help improve our mood throughout the day by emitting a low-intensity current to our brain, which increases the secretion of specific enzymes as shown by Dogrucu et al (2020). This is especially useful to help people with Depression or illnesses such as Schizophrenia. The effect of the current and the subsequent secretion of the enzyme is also recorded and accessible by the concerned doctor.

### 2.1.3 Conclusion and Challenges to Medical IoT

Security is a major issue in Medical IoT due to the sensitive nature of the data, especially since most of the critical health monitoring applications employ a three-layer architecture as observed from Table 1. Although, multiple encryptions exist when the data is transferred from the sensor to the remote cloud server as demonstrated by Haghi et al (2017), security challenges are still predominant in IoT which include confidentiality, authentication, access control, privacy, trust and policy enforcement Balte et al (2015).

Table 1: Architecture and Applications of IoT in Healthcare

| Architecture | Applications | Articles |
|---|---|---|
| **Three-Layer** | Predictive Health Monitoring | McDuff et al (2014), Bergenstal et al (2010), Paul et al (2024), Ramkumar et al (2023), Tuckson et al (2017) |
| | Remote Patient Monitoring and Diagnosis | Adib et al (2015), Hasler (2014), Dias and Paulo Silva Cunha (2018), Dogrucu et al (2020) |
| **Four-Layer** | Remote Health Monitoring from Motion Tracking Sensors | Antonio et al (2021), Mdhaffar et al (2017), Fangyu et al (2021), Zhang et al (2019) |
| | IoT-based UAVs for miscellaneous medical applications | Kumar et al (2019), Ackerman and Strickland (2018), Krishna et al (2018), Sanjana and Prathilothamai (2020), Rosamond et al (2020) |

Another facet of security, that is often overlooked, is the ability of the ML or DL algorithm to be immune to adversarial attacks, especially in medical image segmentation tasks Shukla et al (2024). There have been many attempts to tackle this issue by fine-tuning the DL models using techniques such as transfer learning Alharbi (2024) or adapting a two-stage learning mechanism, with one stage being training on the actual dataset and second stage being training on the anomalous dataset Zafar (2024).

Secondly, the biosensors utilized in wearable devices require specific on-body placement positions or body postures to provide reliable measurements Baig et al (2019). The used of smart clothing, another kind of wearable, is often limited by its application in a variety of climatic and altitude conditions. Due to these variable conditions the body experiences natural changes like increased heart rate and perspiration that peak signal responses and lead to incorrect data Chen et al (2016).

Other challenges, more apt in the case of the wearable devices that incorporate IoT, include wearability and power consumption. As the device itself needs to be worn for extended periods of time it is advisable to make it comfortable and easy to wear. Additionally, all IoT devices need power to work. Therefore, it is necessary to consider the power requirements to reduce user concern Haghi et al (2017).

### 2.1.4 Future Scope for Medical IoT

The recent advancements in health monitoring IoT have been extensively studied above. In addition to this, there have been several AI-based prediction models for health diagnosis as mentioned in Shukla et al (2024); Alharbi (2024); Zafar (2024). The combination of these models, with the latest Generative Models such as GPTs, can help provide more accurate diagnosis for simpler problems as studied by Wang et al (2024b). These models, in combination with robots, have been utilized for tasks such as nursing services Shojaei (2024). Further developments in this field can lead to the development of human-like doctor robots, for a variety of applications.

## 2.2 Agriculture

Agriculture has been a basic and most employed field especially in Asian countries where approximately 70 percent of the population depend on agriculture for their livelihood. In the last 20 years, automation and IoT are slowly making their way into the agriculture field and improving both the quality and quantity of the yield. Some of the applications are Precision Farming, Agricultural Drones, Smart Greenhouses, and Livestock Monitoring.

### 2.2.1 Precision Farming

Precision farming, also called site-specific agriculture, involves the use of technology to record and monitor the response of crops to inputs, like fertilizers, and thus prevents surplus inputs from being supplied, increasing the sustainability of agriculture Webber et al (2017). Often, precision agriculture is confused with smart farming, but there is a major difference between them. Precision agriculture only considers the field's variability, whereas smart farming provides a more exhaustive analysis and performs precise actions like decision-support information and task automation Wolfert et al (2014, 2017). In other words, precision agriculture only gives the data, and smart farming can also perform tasks based on the data. There are many factors that affect the agricultural yield, like climate conditions, moisture and temperature of soil, use of fertilizers, and water monitoring. An IoT network receives data about each of them and suggests an action. The data is collected with the help of various IT components like sensors, robotics, automation, control systems, variable rate technology, etc. The data collected is then stored in the cloud and can be accessed via any mobile device depending on the programming of the network. Prasetya et al (2022) attempted to integrate water monitoring, monitoring the truck fleet where the produce arrived, and fire management. Water monitoring involves the monitoring of water and pH levels. Truck fleet monitoring can be used to identify the factors that affect the seeds and other raw materials that come to the farmers and make the necessary modifications. Fire management is used to analyze factors affecting the spreading of fire like wind direction and concentration of heat in particular areas.

Fig 8 shows the architecture of an IoT-based precision farming network. This method of farming has improved production in numerous instances, as mentioned in Bhanu et al (2014); Rekha et al (2017); Chen et al (2019); Bazargani and Deemyad (2024); Sharma et al (2024); Vatin et al (2024).

In order to assist the data acquisition process for precision agriculture, a common technique used has been the employment of drones. According to Ahmed et al (2018), the drones used in agriculture may be autonomous, single or shared remote control, or mobile controlled, which is a combination of autonomous and shared control established via mobile devices. Autonomous drones are drones that don't need any control mechanism like remotes, and they would be fed with flight plans for their operation. On the other hand, Controlled drones are drones that are remotely controlled by a control system established via an IoT network. Another method of precision agriculture is using an enhanced Augmented Reality (AR) based IoT network. Phupattanasilp and Tong (2019) and Ponnusamy and Natarajan (2021) have used such a mechanism in implementing precision farming. Though both aim at crop monitoring and improving the quality of agriculture, the latter paper combined various concepts like Machine Learning and used drones with cameras for camera monitoring, whereas the former attempted to create such a network with the use of stationary cameras. Although implementing the methods in Ponnusamy and Natarajan (2021) is a little tedious compared to Phupattanasilp and Tong (2019), the mobility of drones aids in covering the entire field, as compared to the use of stationary cameras.

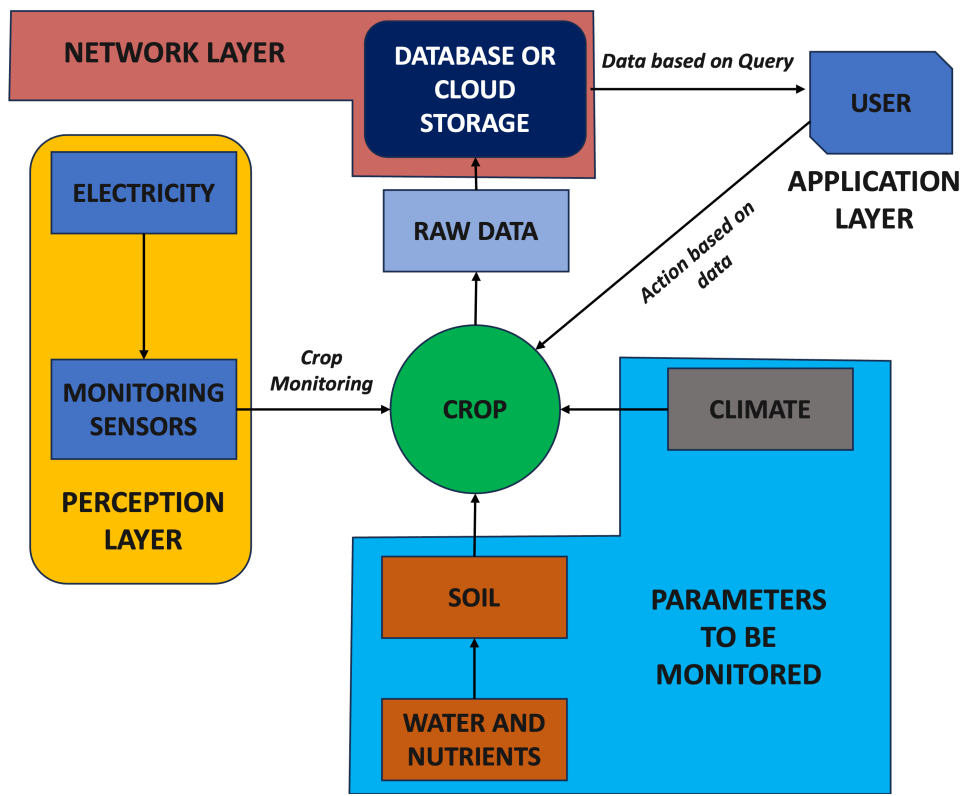

**Fig. 8**: Operational Flow of Precision Farming Balaji (2020)

Nachankar et al (2018) suggested a method of prescribing the water and fertilizer requirement based on the crop and NPK levels, Nitrogen (N), Phosphorus (P), and Potassium (K), of the soil. Initially, the user selects the crop and its growth stage. Then, the area of cultivation is entered in m2. The network calculates the water requirement and displays the value in litres. Then, the NPK values are selected, which helps select the fertilizers suitable for the fields. Saha et al (2018), Bhuvaneshwari et al (2021), and Park et al (2018) have all developed IoT drones that can be employed in crop and forestry monitoring.

Agricultural drones can also be employed in Precision Farming as illustrated above with multiple sensors like pH sensor, soil moisture sensor and temperature sensor in order to retrieve more data and improve the decision-making support system by suggesting choices with more data in hand. They can also be used for irrigation and to spray fertilizers on the field. The health of the crops is also assessed by use of visible and near-infrared light to identify which plants are affected by fungi or bacteria by examining the levels of deviation of reflection of green and the infrared lights Mazur (2016).

Like every other field, AI has also found its way into precision farming as well. One of the most popular applications of AI in this field is for weed and pest detection based on semantic segmentation of image data acquired using drones Singh et al (2016). In addition to this, the use of neural network-based control of water flow for irrigation purposes has also gained significant popularity among researchers as observed in Kashyap et al (2021); Santosh et al (2024); Kaur et al (2024).

### 2.2.2 Farming in Smart Greenhouses

Greenhouses are glass structures with transparent walls and roofs inside which plants that require regulated climatic conditions are grown. IoT can be employed here by controlling factors such as temperature, light, humidity, watering, etc. from a mobile device. Without IoT, these factors might have to be manually controlled from the greenhouse itself. We can also make these factors automatically vary by receiving the data about the plant grown in the greenhouse and the factory settings that need configuring for that plant.

The architecture of an IoT system for greenhouse monitoring and control is shown in Fig 9. The data is acquired with the help of the required sensors (such as temperature, humidity, pH) and then transferred to the Acquisition Node. These two together constitute the Data Acquisition Unit. This data is then transferred to the User via the Gateway and Server. Then, the command from the user is received and transmitted to the Control Unit via the same path and the necessary commands are executed.Wang et al (2020)

Kumar et al (2022) has employed such a method with the help of Global System for Mobile Communication (GSM), Arduino, and other real–time sensors to sense the data, which are all powered by solar power. Then, the data is transmitted and can be viewed and modified with the help of any mobile device that is connected to the internet. Another major problem in greenhouses is the formation of frost on the leaves due to the freezing of the water. This can help in the depletion of the amount of water and food reaching various parts of the plant and, thus, can adversely affect the health and growth rates of the plant. Castañeda-Miranda and Castaño-Meneses (2020a,b) proposes two methods using IoT to monitor and control the frost formation on the surface of the leaves. Castañeda-Miranda and Castaño-Meneses (2020a) employs a combination of IoT, fuzzy, and artificial neural networks are used for monitoring the amount of frost formation, whereas, the anti-frost irrigation system utilized in Castañeda-Miranda and Castaño-Meneses (2020b) is controlled by interaction with a website with the help of GSM and internet services.

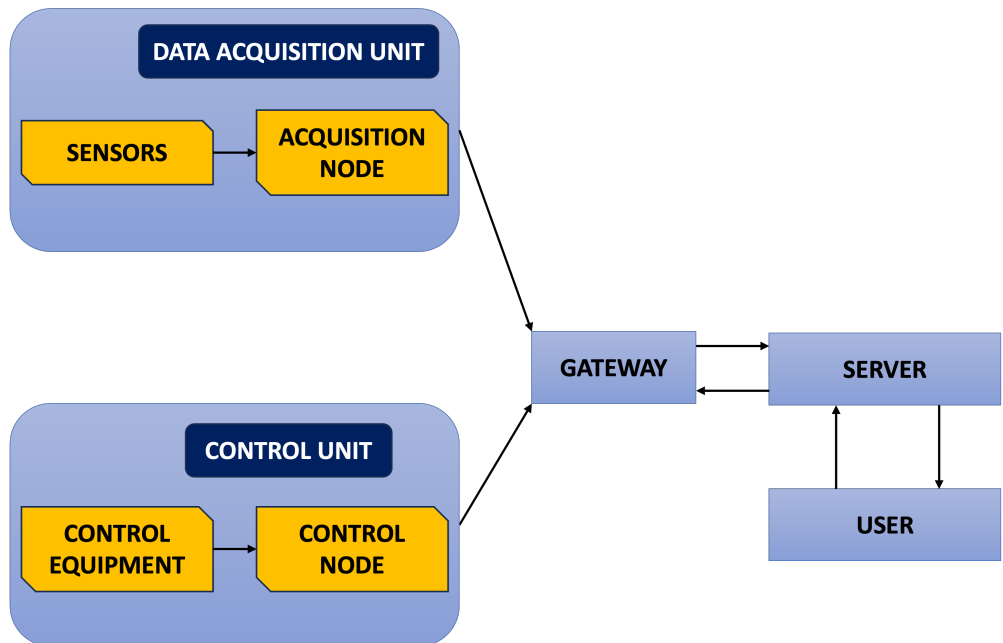

**Fig. 9**: Architecture of IoT-Controlled Greenhouse

### 2.2.3 Livestock Monitoring in Dairy and Poultry Farms

According to the Food and Agriculture Organization of the United Nations (2019), approximately 750 million people are in the dairy farming industry. With the help of technological advancements, such a vast industry is further expanding and making significant progress. The IoT and Artificial Intelligence (AI) methods combined have helped dairy and poultry farmers monitor the conditions of their cattle more frequently and thus improve production by taking action within a short period. For instance, Iwasaki et al (2019) developed a system with biological sensors attached to every cattle. This way, their health conditions like temperature, blood flow, etc, could be monitored. With the help of an IoT network, this information is stored in the cloud, and if any deviation is detected, the respective owner or farmer is alerted. He also suggests that if there is an anomaly in the temperature, the air conditioning or fan can accordingly be varied automatically with the help of coupling with AI mechanisms.

Alonso et al (2020), on the other hand, employs a technology called Global Edge Computing Architecture (GECA) as mentioned in Sittón-Candanedo et al (2019). There are three layers to this kind of architecture: IoT, Edge, and Business Solution Layers, as shown in Fig 8. The IoT layer is the physical layer that consists of hardware components like monitors, sensors, actuators, and nodes. This establishes a connection with the Edge Layer through wireless technologies such as Wi-Fi, Bluetooth Low Energy (BLE), ZigBee, etc, and transmits the raw data sensed by the sensors. The Edge Layer is the central layer, which receives the data from the IoT layer, filters it, and pre-processes it before sending it to the next layer, the Cloud or Business Solution Layer via the Edge Gateway. This also acts as a security or a firewall, as not all data that is received are transmitted; only a certain amount of data is transmitted after filtering. The Business layer performs analytics and cloud management depending on the application. The access control also lies in this layer, and it is up to the user to make the data available with or without an authentication key (private or public). Alonso et al (2020) also proposes a technology called SmartDairyTracer, which uses GECA to monitor livestock farms, used to feed the livestock, real-time location sharing, RFID and QR scanners for packaged goods, and the condition of the product during transportation. It is a strap attached to the cattle so that it is in contact with the skin of the cattle. This is essential to monitor the biological conditions of the cattle. The tracer also relays the live location.

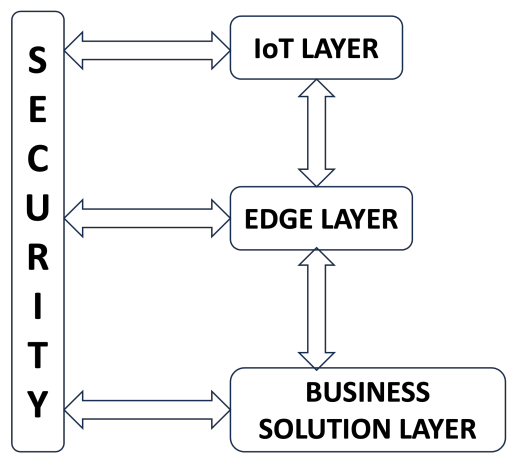

**Fig. 10**: Global Edge Computing Architecture

### 2.2.4 Conclusion and Challenges of IoT in Agriculture

Though IoT has led to a significant advancement in the agriculture industry, it has its own shortcomings. Firstly, like any other device, the machines can't be entirely relied upon as they don't always function perfectly. The safety of sensors and communicating devices in agricultural fields against natural calamities and theft is a major concern for their reliability. In livestock monitoring, there are various challenges, like the sensor mentioned in Iwasaki et al (2019) has to be in direct contact with the skin, which cannot be the case as the livestock may have hair or fur preventing contact. There may also be breakage when a large scale of livestock is being transported from one place to another.

Table 2: Architecture and Applications of IoT in Agriculture

| Architecture | Applications | Articles |
|---|---|---|
| **Three-Layer** | Crop Monioring (using stationary sensors) | Balaji (2020), Ponnusamy and Natarajan (2021), Nachankar et al (2018) |
| | Precision Farming and Irrigation | Mazur (2016), Kashyap et al (2021), Santosh et al (2024), Kaur et al (2024) |
| | Livestock Monitoring in Dairy and Poultry Farms | Iwasaki et al (2019) |
| **Four-Layer** | Crop Monitoring (using drones) | Phupattanasilp and Tong (2019), Ahmed et al (2018), Saha et al (2018), Bhuvaneshwari et al (2021), Park et al (2018), Singh et al (2016), Soman and et al. (2024) |
| | Smart Farming in Greenhouses | Wang et al (2020), Kumar et al (2022), Castañeda-Miranda and Castaño-Meneses (2020a), Castañeda-Miranda and Castaño-Meneses (2020b) |

In remote areas, where farming is practiced, obtaining an internet connection might be a little challenging. Receiving a large amount of heterogeneous data and analyzing and establishing a decision-support mechanism will be tedious. This might also result in the occurrence of unforeseen and unexpected errors and will affect the yield. Any IoT network entails the risk of hacking and, thus, loss of privacy and data. Also, more power is required to run all the hardware involved in such a network.

In addition to this, Table 2 illustrates that the existing architectures work satisfactorily for their corresponding applications. However, one major issue is that, control applications such as irrigation, might require a four-layer architecture (i.e.) addition of a security layer, in lieu of the currently popularly employed three-layer architecture. Secondly, for mobile applications such as monitoring using drones, the raw data acquired maybe noisy as illustrated in Yablokova and et al. (2024). Although there have been multiple DL-based methods developed for improving the quality of data as shown by Ahmed and et al. (2024); Mamatov and et al. (2024), it may require the addition of another layer for data processing, which might increase the latency.

### 2.2.5 Future Scope of IoT in Agriculture

With the rising popularity of GPTs and robots in every field, this field will also see the role of IoT-connected devices integrated with these technologies to enhance their usability. A major part of this would be moving towards "Agriculture 5.0", in which there is significant amount of automation based on AI and robotics Holzinger et al (2024). In addition to this, the further fine-tuning of the existing machine learning models can aid in enhanced performance of these models in real-world agricultural applications, as illustrated by Mohyuddin et al (2024), who compared various ML models used for smart farming and evaluated their performing capabilities. We observe that the increase in training size can improve the performance of the ML models.

# 3 Impact of IoT in Safety

## 3.1 Fire Fighting

Thousands of fire accidents occur every year across the globe. These accidents lead to the loss of lives of many firefighters. Technologies like IoT and UAV (Unmanned aerial vehicles) have been utilized in detecting the fire and extinguishing it quickly and without much damage to the firefighters.

### 3.1.1 Detection of Fires

The traditional technologies are inaccurate and precise enough to make well-informed decisions. The IoT system fitted with several devices like infrared sensors, visual cameras and other embedded sensors helps us detect and monitor the spread of the fire. Sarobin et al (2018) uses an IoT-enabled drone for detecting the fire, which boosts up the fire detection accuracy and thus helps in saving many lives. Kanwar and Agilandeeswari (2018) used an IoT based firefighting robot which alerts the user when fire occurs and also starts extinguishing the fire with the help of $CO_2$ and water pumps. Vijayalakshmi and Muruganand (2017) proposed a layered structure for an IoT based fire monitoring system. This layered structure consists of sensing, transport, service and application layer for monitoring the fire as shown in Table 3. In the sensing layer the data generated by the infrared (Flame) sensors and other embedded sensors are collected. In the transport layer, these data are transported to different networks and platforms through Internet, LAN, mobile network, etc. The service layer involves storing, integrating and visualizing the generated data. The application layer is used for sharing information to the concerned authorities for making decisions. Sarobin et al (2018) uses a drone equipped with an Arduino UNO board, flame sensor, inbuilt camera and a WIFI module for detecting the fire and transmitting the information to a cloud platform like ThingSpeak. Kanwar and Agilandeeswari (2018) used an IoT based firefighting robot fitted with a flame sensor, gas sensor, MHZ-14 $CO_2$ sensor and camera. As soon as the flame sensor detects a fire, the robot sends an alert to the user via cloud which can be viewed easily with the help of an android application installed in the user's device. The user can then view the live video stream of the fire generated by the inbuilt camera with the help of the application. This helps in detecting the hotspots and the direction of spread of fire which is important for taking further decisions. Table 4 shows the architecture used for different fire fighting applications. Even though three layer architecture is simple and easy to implement, it has security concerns and hence four layer IoT architecture is preferred for most of the applications.

Table 3: General IoT System Framework

| Perception Layer | Network Layer | Service Layer | Application Layer |
| --- | --- | --- | --- |
| Firefighting Facility | Police Special Network | Heterogeneous Integration | Firefighting Facility Monitoring System |
| Firefighting Equipment | Field Bus | Heterogeneous Integration | Hazard Source Monitoring and Warning System |
| Personnel, Event, Process | LAN | Visualize Service | Firefighting Equipment and Materials Monitoring System |
| Personnel, Event, Process | Internet, WiFi | Data Storage, Interface | Family, Social Units, Fire Brigade, Government, Intermediaries, Manufacturers, Service Providers |

### 3.1.2 Fire Extinguishing

IoT systems are not only capable of detecting the fire and alerting the safety authorities, it is also capable of extinguishing the fire. There are different ways in which the IoT enabled drones and robots can extinguish the fire.

1. **Pumping System:** Mangayarkarasi (2018) designed a firefighting RF controlled robot with a water tank and a pump. The robot senses the fire with the help of a thermistor. The sensor data is then sent to the microcontroller which processes the information. The water is then pumped from the water tanker onto the fire and extinguishes it. This process is done with the help of wireless communication and RF module application. Kanwar and Agilandeeswari (2018) used an IoT enabled firefighting robot capable of distinguishing the fire types. The robot operates different pumps based on the fire types. The firefighting robot sends a signal to the user via cloud as soon as it detects the fire with the help of the flame sensor. The robot also sends the data generated by the carbon dioxide sensor. This data is used to distinguish between different fire types. If the value of MHZ14 $CO_2$ sensor is low, the robot is instructed to operate the $CO_2$ pump and if the value is high, the water pump is operated. This process flow is shown in Fig 11.

2. **Fire Extinguisher Balls:** Jayapandian (2019) used an IoT drone equipped with GPS sensor, camera, collision avoidance sensor and fire extinguisher balls. The drone after detecting the fire sends the data to the cloud which is then accessed by the drone flight planning unit to take further decisions on the flight plan for the drone. The drone, after getting launched, sends its position through a GPS sensor continuously. This data is used by the personnel to drop the extinguishing ball exactly at the site of fire to extinguish it . The extinguisher balls are made of non-toxic environmentally friendly materials. The ball gets activated within 15 seconds in contact with the fire. The ball will explode and will release all the non-toxic material it is made of and will extinguish the fire. Multiple drones based on cooperative control can also be used for extinguishing the fire as shown in Ghamry and Zhang (2016). The working of this system is illustrated in Fig 12.

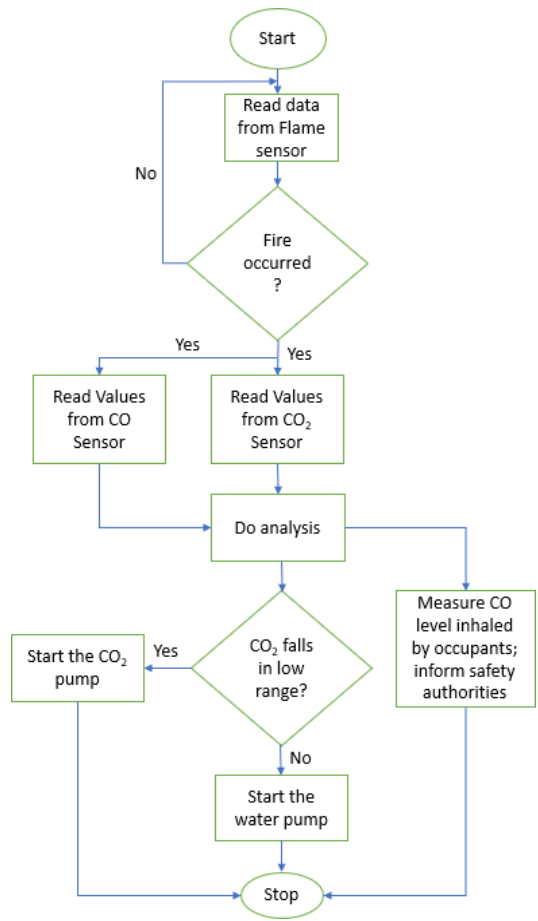

**Fig. 11**: Process Flow for IoT-based Firefighting Robot Kanwar and Agilandeeswari (2018)

### 3.1.3 Conclusion and Challenges in using IoT in Firefighting

Security is a major problem in IoT systems. The communication channels between different devices in the system can be hacked to generate false alarms. Also, these systems could be attacked to hide the existence of real fire and threaten the life of the people involved Toledo-Castro et al (2018). There are two main types of cyber-attacks namely Denial of service (DoS) and Distributed Denial of Service (DDoS). DoS is a cyber-attack where the hacker drains the resources of the real user by sending bulk messages or malicious code. If this attack is done with the help of several nodes, then it is called DDoS. This in turn will make the device perform poorly which may lead to problems Razzaq et al (2017); Džaferović et al (2020).

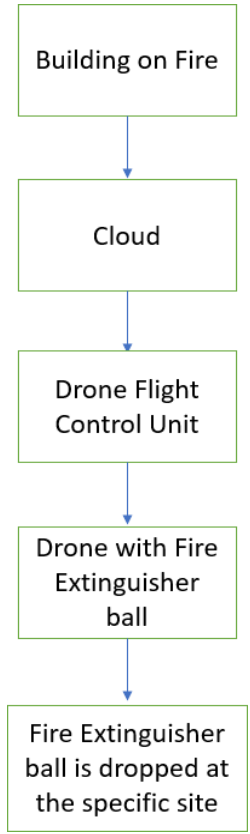

**Fig. 12**: IoT-based firefighting drone

**Table 4**: Architecture and Applications of IoT in Fire Fighting

| Architecture | Applications | Articles |
| --- | --- | --- |
| **Three-Layer** | Fire detection in hostile environments | Zhang et al (2023); Bhattacharjee et al (2012) |
| **Four-Layer** | Hazard and Firefighting equipment Monitoring System | Umar et al (2020), Sarobin et al (2018); Kanwar and Agilandeeswari (2018); Mangayarkarasi (2018) |
|  | Community alert system | Vijayalakshmi and Muruganand (2017) |

### 3.1.4 Future Scope of IoT in Firefighting

The field of Virtual Reality has gained significant popularity in the last 5 decades. This technology provides a platform for development, testing and training for personnel as well as robots as illustrated by Raman and Vyakaranam (2024); Asha et al (2024). The flexibility and scope to bridge the "real-2-simulation" gap, makes it a field with high potential for development.

Secondly, if we consider firefighting UAVs in particular, they still possess the issue of instability of the UAVs employed. This directly affects the accuracy of data acquired, which directly impacts the functionality of the UAV itself Wang et al (2024a). The solution to this problem will make UAV a reliable robot in different applications, which are analyzed in this article.

## 3.2 Women Safety

### 3.2.1 Safety-Trigger Devices

According to the World Health Organization (WHO) (WHO) one out of three Women have been victims of physical/sexual violence. With the ever-increasing cases of harassment against women, most of which goes unreported and some of which even end in rape and murder - the use of technology to design a security system for women so that they don't feel helpless while facing such social challenges, can be godsend. It is possible to take preventive measures to ensure the safety of women using systems with IOT, locate the person with GPS and call for help using GSM.These safety devices can be incorporated in mobile phones, wearables etc by which alerts can be sent to the stored emergency contacts or to the authorities. The general architecture of these safety alert systems is the three-layer architecture as illustrated in Fig 13. The working principles of some of the safety devices utilizing this architecture are mentioned below.

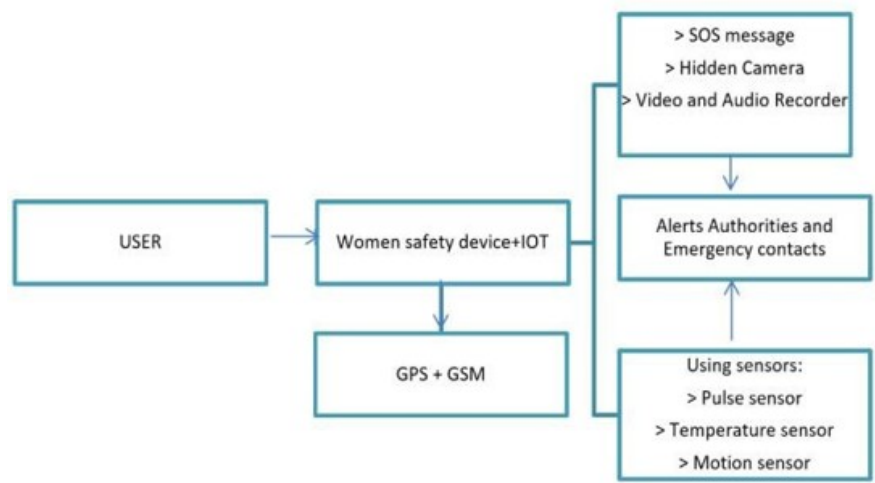

**Fig. 13**: General Architecture of Women Safety Systems

1. **Panic Button Trigger Devices:** Advanced RISC (Reduced Instruction Set Computing) Machines (ARM) is a processor that can be used in electronic devices such as smartphones and wearables. It has a lesser number of circuits and therefore can be used in small devices. The processor requires a minimal number of instructions and consumes less power. As the name suggests these devices get triggered as soon as the user presses a button Zyla et al (2024). Monisha et al (2016) developed a device called "FEMME" constructed with an ARM controller connected to an application via Bluetooth. The system records audio, identifies hidden cameras, sends SOS messages and makes calls to the authorities and emergency contacts. Hanif et al (2020) uses an Arduino as its microcontroller and on pressing the panic button, it sends live location and calls the law enforcement through the SIM800L module. It has a speaker and microphone to communicate with pre-saved contacts. Akram et al (2019) in his work proposed to use fingerprint as a trigger, an android application to suggest possible safe locations for the victim and also a shockwave generator to attack the perpetrator. Nagarji et al (2024) has also developed a similar system, which combines the capabilities of a teaser device and ESP32 module to trasmit the current location via GPS to a centralized base.

2. **Drones:** Drones could be triggered either upon receiving a signal from a panic button on a wrist band or an application. Umar et al (2020) designed a drone which gets triggered upon receiving a signal and communicates it to the nearest police station or activates an alarm or releases tear gas pellets. The drone also sends videos and pictures of the crime scene to the police station. A drone is of advantage, as it can help the person at that particular moment because of the loud buzzer - that can grab the attention of people nearby Cooney et al (2024).

3. **Smart Band / Smart Wearables:** These devices are carried or worn on the wrist. They are triggered by either pressing a button or by speech recognition of a keyword or by detection of variations in the parameters measured by sensors. On receiving a trigger signal, the device with GPS and GSM sends location and alerts to authorities and emergency contacts. Harikiran et al (2016) and Budebhai (2018) employ the use of heartbeat sensor and temperature sensor but do not consider the variation in body temperature and heart beat pattern from person to person. By including Machine learning Muskan et al (2018) collected data from the temperature and heart beat sensor and an alarm was triggered based on the threshold of the collected data. In a more recent study, Verma et al (2023) developed a smart band for risk alert using a three-layer architecture.

### 3.2.2 Conclusion and Challenges of IoT in Women Safety

The proposed devices and systems make use of smartphones and modern technology which might be a challenge in terms of resistance to use technology, internet connectivity and cost. Moreover, majority of the aforementioned devices utilize the three-layer architecture, with the perception layer being the trigger button and transmission layer be a GPS module. However, this mode of trigger activation might fail in cases with less network connectivity or if the attacker possesses a device a GPS Jammer as mentioned in Mitch et al (2011). In addition to this, based on the above survey, especially using the devices mentioned under the Smart Band and Panic Trigger Devices sections of Table 5, it is inferred that the effectiveness of these devices highly depends on the time it takes the signal to alert the police stations and also the proximity to the policemen which would be a challenge in case of a remote location. Therefore, complete safety is not assured as the attacker could cause harm before the arrival of help.

Table 5: Architecture and Applications of IoT in Women Safety

| Architecture | Applications | Articles |
|---|---|---|
| **Three-Layer** | Panic Trigger Devices | Zyla et al (2024), Monisha et al (2016), Hanif et al (2020), Akram et al (2019), Nagarji et al (2024) |
| | Smart Watches/Wearables | Harikiran et al (2016), Budebhai (2018), Muskan et al (2018), Verma et al (2023) |
| **Four-Layer** | Monitoring Drones | Umar et al (2020), Cooney et al (2024) |

### 3.2.3 Future Scope of IoT in Women Safety

There have been many studies conducted to alleviate the aforementioned challenges. However, in all these studies, one fact that still is not properly addressed is how the device will perform when the network is not present. Farooq et al (2023) proposes a three-layer architecture, with multiple options for transmission layer for transporting data based on the output from the ML-based danger detection algorithm. This ensures that the data is transmitted one way or the other. However, a reliable solution is yet to be found to address this issue.

## 4 Surveillance and Mapping

### 4.1 Mapping for Navigation using IoT

### 4.1.1 Mapping and Navigation for UAVs

All the applications that have been discussed until now in the previous sections have required the movement of a device aided by IoT (predominantly, UAVs) from its docking position to a specified target location. In order to execute this in real life, a process called Mapping is used to help the UAV understand the world around it.

    The UAVs can be operated in two modes: Supervised and Unsupervised. In the supervised mode of operation, the map is generated in advance by techniques discussed below and the UAV simply connects to a cloud server to read this data in order to avoid obstacles. Unsupervised method of operation requires the use of an intelligent UAV that can map out its own immediate surroundings even when already in flight. This paper does not discuss the Unsupervised mode of operation in detail as the technology required for its use is very limited and expensive and is not a good investment in applications such as the one discussed above.

A very integral part of trying to plan the motion of a robot or a UAV in a space in Supervised mode of operation is the actual 2D or 3D spatial map of the place Burigat et al (2017). This is applicable irrespective of the final field of application of the UAV. Despite the supervised nature of operation, the preloaded maps alone are often insufficient. Real time obstacle identification is an integral part of any UAV. This is especially true in disaster management or situations where change has to be monitored over time Früh (2004).

In order to generate real-time and accurate maps, several advanced mapping techniques using IoT have been utilized, especially in swarm-base control for scene-reconstruction and surveillance applications as illustrated in Machaiah and Akshay (2019); Garg et al (2024). All the data is pooled in a cloud server and the individual maps are superimposed to generate a live map of a region. Although IoT itself has no role in generating the first map of a place, it has definitely aided in continuous monitoring and updating of maps in real time. This can be used to determine traffic conditions and intimate road users on roads under repair or maintenance Kheder and Mohammed (2024). Figure 14 shows the general architecture of an IoT-based map generation application Hakiri et al (2015).

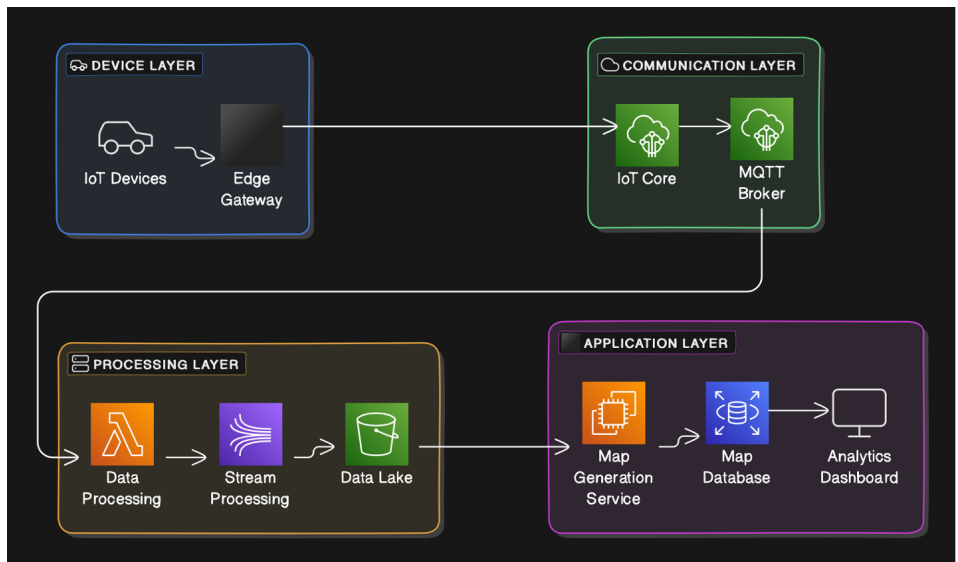

**Fig. 14**: IoT-based Map Generator

Recent advancements in the field of robotics and UAVs, has seen the application of these maps for safe navigation through the environment. For performing this mode of operation in UAVs, some of the commonly utilized methods are aerial Simultaneous Localization and Mapping (Aerial-SLAM) Milford et al (2011), aerial-ground cooperation and autonomous navigation Chai et al (2024) and visual planar semantic SLAM (VPS-SLAM) Bavle et al (2020). These algorithms not only aid in mapping the perceptible surroundings of the UAVs, but also contain architectural nodes for incorporating control signals for performing safe navigation through the surroundings. The architecture of a generalized SLAM framework using IoT is illustrated in the Figure 15 Chai et al (2024). The SLAM algorithm in this diagram will be one of the algorithms mentioned above based on the sensors on-board.

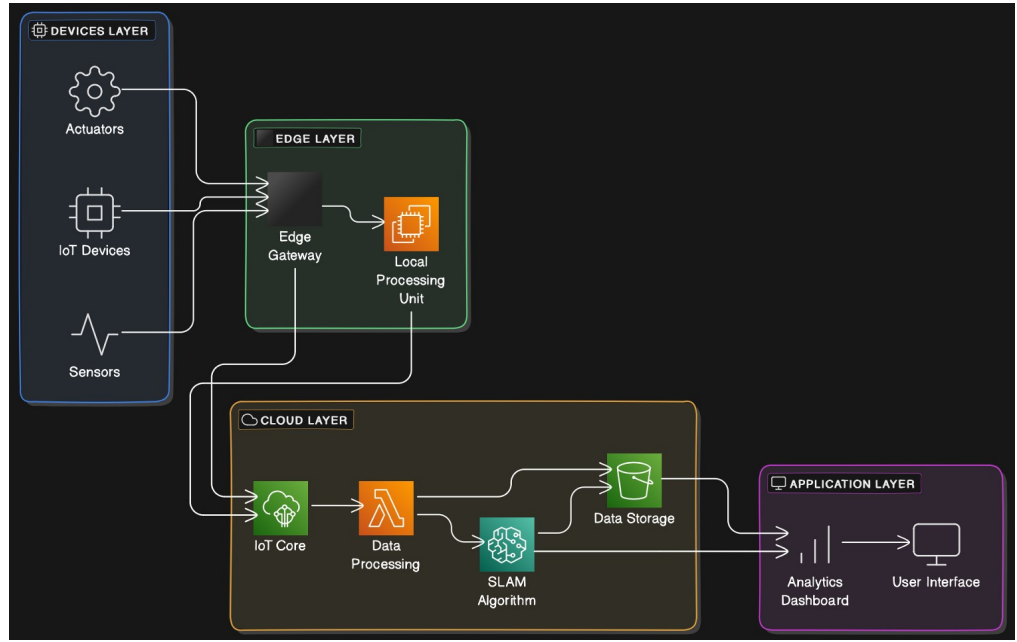

**Fig. 15**: SLAM-based robot navigator using IoT

### 4.1.2 Mapping and Navigation for Autonomous Vehicles

In autonomous vehicles, on the other hand, IoT has played a major role in providing accurate maps for navigation through complex and dynamic environments such as busy streets Biswas and Wang (2023). However, in order to establish a robust working autonomous vehicle, in addition to perception, fast decision-making with minimal latency is essential. Balachander and Venkatesan (2021) accomplished this enhancing the communication capabilities between the network and application layer by using a combination of an ATMega328P microcontroller for storing the digital data and an ESP8266 for transmission of the data to the cloud server. Thus, it is observed that they have employed a four-layer architecture, with data storage being performed at 2 locations, cloud and the microcontroller itself. In another study conducted by Hossai et al (2017), a similar architecture was employed where the image data obtained from a Raspberry Pi camera was used as the input to control the motors of an autonomous vehicle. Thus, it can be inferred that for this application, four-layer architecture is mainly utilized. Some of the developed autonomous vehicles have been extended to commercial applications such as autonomous taxis Vaidya and Mouftah (2020). The IoT-based swarm control of autonomous vehicles is termed as Internet of Vehicles Contreras-Castillo et al (2017).

Since these are commercial applications, a five-layer architecture is employed to determine the business model based on the data that has been obtained from the previous applications. Moreover, the fuel efficiency has been calculated for data-driven navigation based on traffic patterns. Based on this data, the sustainability of autonomous vehicles can be significantly enhanced Chen et al (2021). In addition to this, the data from the application layer can also be used to accomplish scheduling tasks such as carpooling as illustrated in Malik et al (2021). With the advancements in cognitive computing, there is a modified five-layer architecture presented by Chen et al (2018), which includes an additional cognition layer in between the communication and application layer. Cognitive computing is the field of research that enables computer to mimic the computing techniques of the human brain, which can lead the computer to learn the psychological aspects of computing Gutierrez-Garcia and López-Neri (2015).

### 4.1.3 Conclusion and Challenges for Mapping using IoT

Table 6: Architecture and Applications of IoT in Mapping and Navigation

| Architecture | Applications | Articles |
| --- | --- | --- |
| **Four-Layer** | Mapping | Burigat et al (2017), Hakiri et al (2015), Machaiah and Akshay (2019), Kheder and Mohammed (2024), Garg et al (2024) |
|  | Autonomous Vehicle and Robot Navigation | Milford et al (2011), Chai et al (2024), Bavle et al (2020), Biswas and Wang (2023), Balachander and Venkatesan (2021), Hossai et al (2017) |
| **Five-Layer** | Autonomous Cooperative Taxis | Vaidya and Mouftah (2020), Malik et al (2021), Chen et al (2018), Chen et al (2021) |

The architectural information of each of these applications is illustrated in Table 6. Although most of the aforementioned applications require additional nodes for data pre-processing to incorporate the mapping and navigation capabilities, the overall architecture of the IoT framework remains the same as seen in Figure 14 and Figure 15. An additional layer is incorporated to perform tasks mentioned in Chen et al (2021) and Malik et al (2021).

### 4.1.4 Future Scope for IoT in Mapping and Navigation

The concept of sensor fusion has gained immense popularity over the past few years. It is the process of combining multi-modal data from various sensors present on the system and then transporting the data acquired. However, there still exists some amount of noise in the data acquired by the sensors, and when this data is combined, accumulation of noise is observed at the application layer Krishnamurthi et al (2020). The ability to tackle this issue has been studied in various applications Ronald and Raman (2024); Liu et al (2024). However, a complete study and validation of these techniques is yet to be performed. The study on this will help enhance the capabilities of existing technologies that utilize multi-modal data for manipulation such as UAVs Chai et al (2024) and autonomous vehicles Kong et al (2024).

## 4.2 Smart Homes

In the past home security meant an alarm going off in response to a burglar breaking in. However, if there is no individual present, this alarm system is not effective. With the incorporation of IOT to home surveillance, it is possible to ensure security, comfort, smart living and energy saving. Table 7 shows some of the devices that can be used in home surveillance and their role. In addition to this, smart homes also possess features for performing energy-saving. Recent studies have also been focused on developing multi-purpose systems, which are aimed at combining both these tasks using modern multi-modal data processing techniques like data fusion. The principles of security and energy-efficient systems are outlined in Section 4.2.1 and multi-purpose systems are explained in Section 4.2.2.

**Table 7**: General IoT System Framework

| S.NO | Smart Home Security Device | Function |
| --- | --- | --- |
| 1. | Motion Sensor | Detects unauthorized movement (motion or vibration) |
| 2. | Biometric Locks | Eliminates the usage of keys and ensures entry only with biometric authentication (fingerprint / facial recognition) |
| 3. | Smart Camera | Provides live feed of the house, night vision, sound detection, etc. |
| 4. | Video Door Entry Systems | Allows the person to control access to their home on-site or remotely |
| 5. | Smoke Sensor | Detects the presence of smoke and alerts the person that a fire has broken out |
| 6. | Smart Switches | Can be used to control several appliances and save electricity by automatically switching the device off when not in use |

### 4.2.1 Surveillance Systems for Smart Homes

Anitha (2017) tested a home security system using an Arduino Uno as the microcontroller, Magnetic Reed Sensor to monitor the status of the device, buzzer for sounding alarm, Wi-Fi module and ESP8266 to connect and communicate via the internet and Blynkapp to send push notifications to the user on receiving input from the Reed sensor. However, the Reed switch has several disadvantages such as overheating, susceptibility to breakage and low durability due to the mechanical nature of the switch.

Anwar and Kishore (2016) has designed a home security system using ARM1176JZF-S microcontroller. The system also includes a camera module for capturing images, PIR (Passive Infrared Sensor) for detecting motion, Relay driver for control of Electromagnetic Door lock and a Loud Speaker for Voice alerts. The system sends alert emails to the authorities via TCP (Transmission Control Protocol)/IP (Internet Protocol). Anvekar and Banakar (2017) has designed a very similar system however the alerts are sent through Telegram as it supports AES (Advanced Encryption Standard) and Triple Data Encryption Standard.Khoje et al (2017) reported a security system using Raspberry Pi and Arduino microcontroller. A GSM module is used for sending alerts. This system employs the use of OpenCV (Open-Source Computer Vision Library), MOG2 algorithm for detecting motion by background subtraction method and human discovery by Haar algorithm. This system is also capable of detecting the number of individuals and is unique as it does not use any sensors.

The use of IoT-based sensors can also be extended to enhance the energy-saving capabilities in a household. Several studies have been conducted in this field of research. Dasappa and Somu (2024) proposed the development of such an energy management framework for smart spaces (EMSS). This method uses data from multiple sensors, such as environmental sensors and cameras, for occupancy detection and regulates the power supplied to energy-consuming appliances based on the acquired information. It uses models such as YOLO and MobileNet for object detection and incorporates a custom developed data fusion pipeline. Similar studies using various other technologies have also been conducted by Khan et al (2024) and Ikram et al (2024). The former employs DL methods to find trends in the data acquired and modifies the settings of appliances accordingly, whereas, the latter employs a meta-heuristic optimization technique with the provision of shiftable load hours, based on the user's schedule.

### 4.2.2 Multi-purpose Surveillance Systems

Home surveillance systems designed by Taryudi and Budi (2018), Kumar et al (2018), Kodali et al (2016) include several additional applications such as room temperature and humidity detection using DHT-22 sensor, rain sensor, fire sensor, LDR sensors to monitor light, LPG gas sensor to detect gas leakage and PIR sensor to detect motion. Kumar et al (2018) designed the system using TI-CC3200 Launchpad board which has both the microcontroller and the WI-FI shield embedded which makes it useful for controlling several appliances with one device. The combination of energy optimizers and intruder detectors has also been studied. Netinant et al (2024) accomplishes this by using voice commands of users for automation and passive infrared sensors (PIR) for intruder detection.

### 4.2.3 Conclusion and Challenges in Home Surveillance

Smart home surveillance systems are vulnerable to cyberattacks and hacking. Cyber-attacks cause breach of personal information from laptops and other devices and utilize cameras to spy on residents. As observed from the Table 8, most of them tackle this issue by adding an additional security layer for encryption. Another drawback of smart home surveillance systems is they are expensive and require skilled personnel for setting up. In addition to this, it is very tedious to acquire compatible devices that utilize similar software and architecture. Lastly, since all devices are connected with each other, in the event of failure of one device, the entire system may fail.

Table 8: Architecture and Applications of IoT in Home Surveillance

| Architecture | Applications | Articles |
| --- | --- | --- |
| **Three-Layer** | Surveillance | Anwar and Kishore (2016), Khoje et al (2017) |
| **Four-Layer** | Surveillance | Anvekar and Banakar (2017) |
| | Energy Optimization | Dasappa and Somu (2024); Khan et al (2024); Ikram et al (2024) |
| **Five-Layer** | Multi-Purpose Surveillance Systems | Taryudi and Budi (2018); Kumar et al (2018); Kodali et al (2016); Netinant et al (2024); Ansari et al (2024) |

### 4.2.4 Future Scope of IoT in Home Surveillance

The future of home surveillance within the realm of the Internet of Things (IoT) is poised to revolutionize smart homes, enhancing security, convenience, and integration. Future IoT-enabled home surveillance systems will likely incorporate advanced AI and machine learning algorithms, enabling real-time threat detection, facial recognition, and behavior analysis Ansari et al (2024). These systems will seamlessly integrate with other smart home devices, creating a cohesive network that can automatically trigger alarms, lock doors, adjust lighting, and notify homeowners and authorities of potential security breaches. Enhanced data encryption and cybersecurity measures will be crucial to protect against hacking and unauthorized access. Additionally, edge computing will reduce latency and ensure quicker response times, while innovations in sensor technology will improve the accuracy and reliability of surveillance equipment. As IoT continues to evolve, smart home systems will also focus on energy efficiency and automation, providing homeowners with greater control over their environment as illustrated by Dasappa and Somu (2024); Khan et al (2024); Ikram et al (2024). Importantly, these advancements will be balanced with a strong emphasis on user privacy, ensuring that these sophisticated systems are not only efficient but also respect personal data and privacy.

# 5 Conclusion and Future Work

The Internet of Things (IoT) has experienced substantial development since its inception, significantly impacting various sectors. IoT has enhanced data accessibility and simplified data sharing. In healthcare, IoT has predominantly been applied in patient monitoring, using a variety of sensors to track vital signs. This rapidly growing field has seen considerable advancements over the past decade. The future of IoT in healthcare is poised to benefit from the integration of robotic systems to perform repetitive and intricate tasks, such as surgical procedures, thereby enhancing efficiency and precision. Furthermore, AI plays a pivotal role in healthcare IoT, enabling the analysis of vast amounts of sensor data to detect anomalies, predict potential health issues, and optimize patient care pathways.

In agriculture, IoT has revolutionized practices through precision agriculture, smart greenhouses, and the utilization of drones. These technologies have streamlined farming operations and facilitated informed decision-making regarding fertilizers, crops, and seeds based on weather conditions. The future of IoT in agriculture will likely hinge on automating repetitive or hazardous tasks, such as the application of pesticides, reducing the need for human intervention and increasing safety. AI algorithms processing data from IoT devices can optimize irrigation, predict crop yields, and manage resources efficiently, contributing to sustainable and efficient farming practices.

IoT applications in firefighting include automated pumping systems and the deployment of drones to monitor thermal patterns and identify casualties. Future advancements in IoT-based firefighting systems will focus on innovative technologies like fire extinguisher balls, which can autonomously control and extinguish fires. AI integrates with IoT sensors to predict fire outbreaks by analyzing environmental data and thermal patterns. Autonomous firefighting robots and drones, guided by AI, can navigate hazardous environments, identify hot spots, and deploy extinguishing agents more effectively.

In the realm of women's safety, IoT devices enable precise location tracking and alert authorities during emergencies. Despite the development of various safety devices, crimes against women persist. To enhance the effectiveness and adoption of such devices, integrating them as default features in mobile phones or wearables is essential. AI enhances the functionality of IoT safety devices by providing real-time analytics, pattern recognition, and predictive alerts. AI can improve the accuracy of location tracking and automate the process of alerting authorities, ensuring a rapid response in emergencies.

IoT devices often require assistance in environmental awareness, utilizing mapping techniques such as photography, lasers, and LiDAR. These devices employ a combination of sensors to create a 3D model of their surroundings before deployment. Conversely, IoT has facilitated the real-time creation of maps by aggregating segments in a cloud server, enabling continuous updates and accurate information dissemination.

AI-driven data processing enhances the accuracy and efficiency of mapping techniques used by IoT devices. AI algorithms can analyze sensor data to create detailed 3D models and update maps in real time, facilitating better decision-making and deployment strategies.

The intersection of privacy and security in IoT applications is delicate and multifaceted. Home surveillance systems provide a sense of security and valuable evidence for law enforcement. However, these systems must be highly secure to prevent privacy breaches. AI enhances home surveillance systems by enabling features such as facial recognition, behavior analysis, and anomaly detection. AI-driven security systems can differentiate between normal and suspicious activities, providing timely alerts while maintaining user privacy through advanced encryption and access control mechanisms.

In conclusion, the synergy between AI and IoT holds significant potential to transform healthcare, agriculture, firefighting, women's safety, and environmental mapping. AI's ability to process and analyze data in real-time will drive future innovations, making IoT systems more intelligent, autonomous, and effective across various applications.

# 6 Competing Interests

On behalf of all the authors, the corresponding author states that there is no conflict of interest.

# 7 Funding Information

No funding was received for this project, hence not applicable.

# 8 Author Contribution

The authors confirm their contribution to the paper as follows:

1. **Anush Lakshman S:** Study conception, design, and contribution to study on agricultural IoT.
2. **Akash S:** Study on Mapping, Surveillance, and health-monitoring devices.
3. **Cynthia J:** Study on Safety applications of IoT.
4. **Gautam R:** Study on applications of IoT in firefighting.
5. **Ebenezer D:** Mentor and draft manuscript preparation

All the authors reviewed the results and approved the final version of the manuscript.

# 9 Data Availability Statement

Not Applicable, as this is a survey article.

# 10 Research Involving Human and/or Animals

Not Applicable

## 11 Informed Consent

Not Applicable as no experiments were performed on humans/animals.